# Dynamically encircling exceptional points: *in situ* control of encircling loops and the role of the starting point

Xu-Lin Zhang,[1,2] Shubo Wang,[1,4] Bo Hou,[1,3] and C. T. Chan[1]

[1]*Department of Physics, The Hong Kong University of Science and Technology, Clear Water Bay, Hong Kong, China*
[2]*State Key Laboratory of Integrated Optoelectronics, College of Electronic Science and Engineering, Jilin University, Changchun, China*
[3]*College of Physics, Optoelectronics and Energy & Collaborative Innovation Center of Suzhou Nano Science and Technology, Soochow University, Suzhou, China*
[4]*Department of Physics, City University of Hong Kong, Kowloon, Hong Kong, China*

The most intriguing properties of non-Hermitian systems are found near the exceptional points (EPs) at which the Hamiltonian matrix becomes defective. Due to the complex topological structure of the energy Riemann surfaces close to an EP and the breakdown of the adiabatic theorem due to non-Hermiticity, the state evolution in non-Hermitian systems is much more complex than that in Hermitian systems. For example, recent experimental work [Doppler *et al*. Nature **537**, 76 (2016)] demonstrated that dynamically encircling an EP can lead to chiral behaviors, i.e., encircling an EP in different directions results in different output states. Here, we propose a coupled ferromagnetic waveguide system that carries two EPs and design an experimental setup in which the trajectory of state evolution can be controlled *in situ* using a tunable external field, allowing us to dynamically encircle zero, one or even two EPs experimentally. The tunability allows us to control the trajectory of encircling in the parameter space, including the size of the encircling loop and the starting/end point. We discovered that whether or not the dynamics is chiral actually depends on the starting point of the loop. In particular, dynamically encircling an EP with a starting point in the parity-time-broken phase results in non-chiral behaviors such that the output state is the same no matter which direction the encircling takes. The proposed system is a useful platform to explore the topology of energy surfaces and the dynamics of state evolution in non-Hermitian systems and will likely find applications in mode switching controlled with external parameters.

## I. INTRODUCTION

Exceptional points (EPs) are degeneracies in non-Hermitian systems [1-4]. Unlike degeneracies in Hermitian systems such as diabolic points (DPs) [5,6] whose eigenvalues but not eigenvectors coalesce, at EPs both the eigenvalues and the eigenvectors coalesce, leading to various counter-intuitive phenomena and fascinating applications such as loss-induced transmission enhancement [7], lasing effects [8-11], unusual beam dynamics [12,13], enhanced sensing [14-16], robust wireless power transfer [17], and others [18-23]. The most intriguing feature of the EP is perhaps its topological structure in the sense that adiabatically encircling an EP can result in an exchange of the eigenstate [24,25], unlike the encircling of a DP in Hermitian systems where the eigenstate would only acquire a geometric phase [5,6]. The so-called state flip achieved by adiabatically encircling an EP is made possible by the degeneracy-induced intersection of complex Riemann sheets [24,25]. This phenomenon has been demonstrated experimentally in microwave cavities [26], exciton-polariton systems [27] and acoustic systems [28], where static measurements of the spectra and eigenmodes successfully revealed the topological structure of EPs. However, the outcome is completely different if an EP is encircled in a dynamical process. In dynamical encircling, the output state has been predicted to be determined solely by the direction of rotation in the parameter space regardless of the input state. Such "chiral behavior" [29] is a manifestation of the breakdown of the adiabatic theorem in non-Hermitian systems in the presence of gain and loss [30,31]. The chiral nature of the dynamics has also been theoretically investigated from the viewpoint of stability loss delay [32] and the Stokes phenomenon of asymptotics [33], and a full analytical model has been proposed for a better understanding [34]. It was not until recently that the dynamical encircling of an EP was realized experimentally in microwave [35] and optomechanical systems [36]. The chiral behavior is expected to have promising applications in asymmetric mode switching [35,37] and on-chip nonreciprocal transmission [38].

Although the dynamical encircling of an EP has been demonstrated both theoretically and experimentally, previous studies focused exclusively on encircling loops with the starting/end point near the parity-time-symmetric (*PT*-symmetric) phase [34-38], where the imaginary parts of the eigenvalues coalesce. But what if the starting point of the dynamical process lies somewhere else in the parameter space? Would the chiral behavior persist if the dynamical encircling starts from a point near the *PT*-broken phase where the real parts of the eigenvalues coalesce? These questions remain open. Furthermore, the dynamical evolution of states in non-Hermitian systems in which non-adiabatic transitions (NATs) may occur due to the breakdown of the adiabatic theorem is of fundamental interest. This area is, however, largely unexplored especially experimentally due to the complexity in system design. The recent pioneering work [35] used a modulated waveguide system to realize EP encircling. The system offers an excellent platform to study the dynamics in non-Hermitian systems as the state evolution and NATs can be understood intuitively from the field profiles in the waveguides. However, the encircling loop in the experiment is fixed once the sample is fabricated, and changing the loop in fact requires fabricating new samples. A new platform on which the encircling loop could be controlled *in situ* using, for example, external parameters is highly desirable.

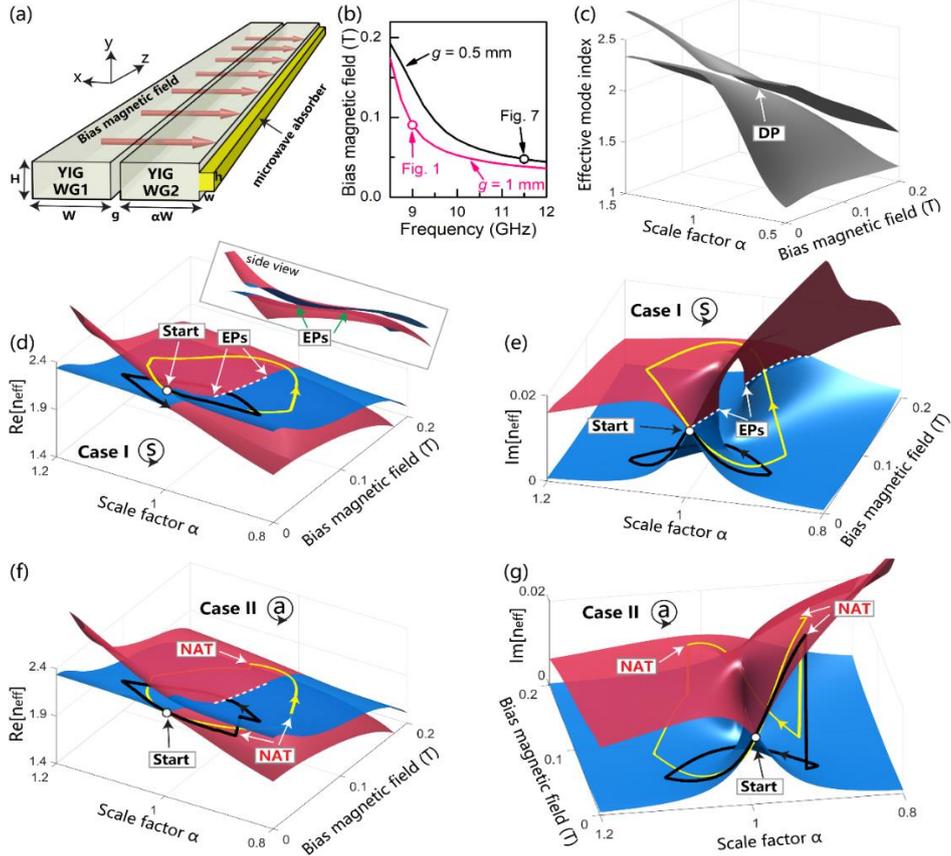

FIG. 1. (a) Schematic diagram of a coupled yttrium iron garnet (YIG) waveguide system with a microwave absorber attached to waveguide-2. A bias magnetic field is applied along the negative *x*-axis. (b) Calculated bias magnetic field at which a diabolic point (DP) emerges in the lossless system as a function of frequency with $g = 0.5$ mm (black line) and 1 mm (red line). Other structural parameters are $W = 8$ mm, $H = 4$ mm, and $\alpha = 1$. (c) Calculated effective mode index as a function of the bias field and scale factor of the lossless system. A DP appears at $B_0 = 0.092$ T and $\alpha = 1$ due to accidental degeneracy. (d)-(e) Calculated real part (d) and imaginary part (e) of the effective mode index as a function of the bias field and scale factor of the lossy system. The two figures show self-intersecting Riemann surfaces with two exceptional points (EPs) at $B_0 = 0.06$ T, $\alpha = 0.988$ and $B_0 = 0.123$ T, $\alpha = 0.982$. The white dashed line in (d)/(e) marks the broken/symmetric phase line. The black and yellow lines represent the trajectory of state evolution for case I with $B_m = 0.08$ T (encircling one EP) and 0.17 T (encircling two EPs), respectively. (f)-(g) Same as panels (d)-(e) except that the trajectories are for case II. In the simulations of (c)-(g), the frequency is 9 GHz and the system parameters are $W = 8$ mm, $H = 4$ mm, $g = 1$ mm, $w = 1.5$ mm, and $h = 2$ mm. The relative permittivity of the absorber is $3+3i$.

In this work, we propose a platform to study the dynamical process in non-Hermitian systems and the dynamical encircling of EPs. On this platform, the trajectory of state evolution in the parameter space can be controlled *in situ* using an external parameter. Our system consists of a pair of ferromagnetic waveguides applied with transverse bias magnetic fields. The waveguide width and the external magnetic field are non-uniform so that when wave scatters through the system, it is effectively traveling along a trajectory in a pre-designed two-variable parameter space, where a pair of EPs with opposite chirality reside. The topological structure of the system can be designed by choosing appropriate system parameters, allowing us not only to dynamically encircle different numbers of EPs (e.g., zero, one or even two) without changing or moving the sample, but also to study the dependence of the dynamics on the starting/end point of the encircling loop. We first realized experimentally the previously discovered chiral transmission behavior [35] by dynamically encircling an EP with the starting point in the symmetric phase. Moreover, our system has two EPs which allow us to dynamically encircle two EPs to reveal the more complex topological structure of energy surfaces. The main finding of this work is that we investigated the dynamical encircling of an EP with the starting point in the broken phase and discovered a non-chiral behavior, indicating that whether the dynamics is chiral or not depends on the starting point. A theoretical model was used to investigate the underlying physics and reveal the role of the starting point.

## II. IN SITU CONTROL OF ENCIRCLING LOOPS WITH AN EXTERNAL FIELD

We start by introducing a platform for studying the dynamical process in non-Hermitian systems. As shown in Fig. 1(a), the system consists of a pair of yttrium iron garnet (YIG) waveguides separated by a small gap. We apply a transverse bias magnetic field along the negative $x$-axis. A microwave absorber is attached to the side of YIG waveguide-2 to introduce asymmetric losses [39] into the system. The background is air. The width of YIG waveguide-2 is controlled by a scale factor $\alpha$, corresponding to a detuning of the system. We first calculated the effective mode index of the waveguide pair system ($W$ = 8 mm, $H$ = 4 mm, $g$ = 1 mm) as a function of the scale factor $\alpha$ and the bias field using COMSOL [40]. In the simulation, the relative permittivity of YIG is set to ~15.26, and the relative permeability tensor of YIG is modeled with a diagonal term $\mu_b = 1 + \omega_m \omega_0 / (\omega_0^2 - \omega^2)$ and off-diagonal terms $\pm i\chi = \pm i\omega_m \omega / (\omega_0^2 - \omega^2)$, where $\omega_m = \mu_0 \gamma_R M$ is determined by the gyromagnetic ratio $\gamma_R$ and the magnetization $M$, and $\omega_0 = \gamma_R B_0$ is determined by the bias magnetic field $B_0$ [41]. The effective mode index is defined as $n_{eff} = \beta_z / k_0$, where $\beta_z$ and $k_0$ are the mode propagation constant and vacuum wave number, respectively. The results for the lossless system (i.e., without the absorber) at 9 GHz are shown in Fig. 1(c). We find that two eigenmodes are supported in the system. A DP emerges ($B_0$ = 0.092 T, $\alpha$ = 1) due to the accidental degeneracy of the two eigenmodes [42]. When the microwave absorber is attached ($w$ = 1.5 mm, $h$ = 2 mm, $\varepsilon$ = 3+3$i$), the effective mode index becomes a complex number, and the DP splits into a pair of EPs [42], exhibiting a self-intersecting Riemann surface as shown in Figs. 1(d) (real part) and 1(e) (imaginary part). The white dashed line in Fig. 1(d) marks the broken phase line on which the real parts of the two eigenvalues coalesce (also refer to the side view). The two end points of this broken phase line are EPs, beyond which are two symmetric phase lines (see the two white dashed lines in Fig. 1(e)) on which the imaginary parts of the two eigenvalues coalesce. The symmetric phase line is a branch cut that connects the lower-loss Riemann sheet (see the blue sheet in Fig. 1(e)) with the higher-loss Riemann sheet (see the red sheet in Fig. 1(e)).

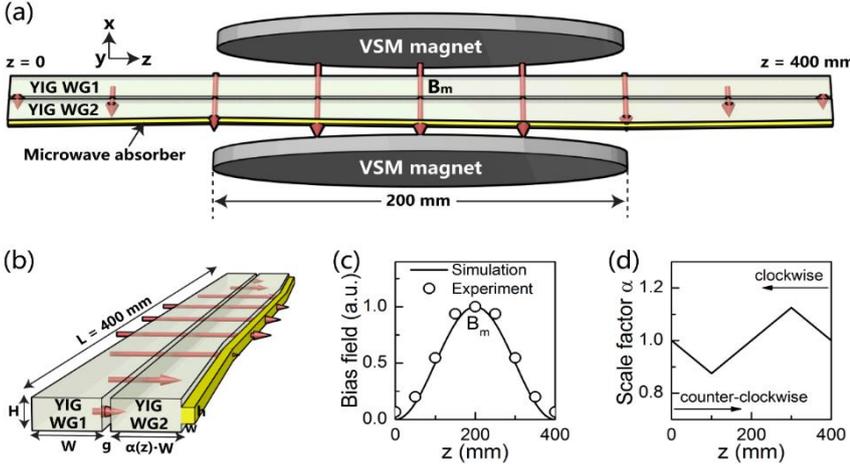

FIG. 2. (a) Schematic diagram of a coupled YIG waveguide system with a length $L$ = 400 mm, where the bias field generated by the two magnets and the width of YIG waveguide-2 vary continuously along the $z$-axis. (b) Side view of the coupled system. (c) Experimentally measured bias field distributions along the $z$-axis (circles), fitted using $B_0(z) = B_m \sin(\pi z / L)$ for numerical simulations (solid line). (d) Variation in the scale factor $\alpha$ along the $z$-axis. The minimum $\alpha$ is 0.875 at $z$ = 100 mm and the maximum is 1.125 at $z$ = 300 mm. Injections from $z$ = 0 and $z$ = 400 mm correspond to counter-clockwise and clockwise loops, respectively.

As we have a Riemann surface containing a pair of EPs, forming an encircling loop requires changing two parameters (the bias field and the scale factor $\alpha$) continuously in space. To implement the encircling, we design a system that is ~400 mm long as shown in Figs. 2(a) (top view) and 2(b) (side view) with the two parameters varying continuously along the waveguiding direction (i.e.,

$z$-axis). The bias field is experimentally generated with a vibrating sample magnetometer (VSM) which has two magnets with a diameter of ~200 mm. The experimentally measured bias field distribution along the $z$-axis is plotted with circles in Fig. 2(c), and $B_m$ denotes the maximum field strength at the center of the waveguides (i.e., $z$ = 200 mm). The field is essentially uniform along the $x$- and $y$-axis in our experimental setup. The field distribution is well fitted using a sinusoidal function (solid line in Fig. 2(c)) for further numerical simulations. The scale factor $\alpha$ is designed to vary along the $z$-axis with a minimum of 0.875 at $z$ = 100 mm and a maximum of 1.125 at $z$ = 300 mm (see Fig. 2(d); also see Supplemental Material for a discussion on the two corners). A two-parameter space is defined in Fig. 3(a), where the locations of the two EPs are also marked. We note that the wave scattering through the system is analogous to a loop in the two-parameter space, with the starting/end point at $B_0$ = 0 and $\alpha$ = 1. Injections from the left ($z$ = 0) and the right side ($z$ = 400 mm) of the waveguide system (see the schematic diagram in Fig. 2(a)) correspond to counter-clockwise and clockwise loops, respectively. Selected examples of the loops are illustrated in Fig. 3(a), where the green, black, and yellow loops are generated at bias field strengths $B_m$ = 0.01 T, 0.08 T, and 0.17 T, corresponding to a dynamical encircling of zero, one, and two EPs, respectively.

The encircling loop in the proposed system can be tuned *in situ* along the $B_0$-axis of the parameter space and the loop size is determined by an adiabatically tunable parameter, $B_m$. Although the loop cannot be tuned along the $\alpha$-axis, such tunability can already enable us to control *in situ* the number of EPs encircled. This was not possible in previous experimental work (see, for example, Ref. [35]), where the encircling loop is fixed once the samples are fabricated. The topological structure of our system is also more complex than previous ones [34-38] due to the presence of two EPs, and the locations of the EPs can be specified by choosing appropriate system parameters. To demonstrate this point, we show in Fig. 1(b) the calculated bias fields required to access the DP in the lossless system as a function of frequency with two different gap distances. The red circle corresponds to the case in Fig. 1(c). When loss is introduced, the DP splits into two EPs and their locations can be specified by choosing appropriate absorbers. Higher-loss absorbers can result in a broader broken phase region whereas lower-loss absorbers can lead to a narrower region

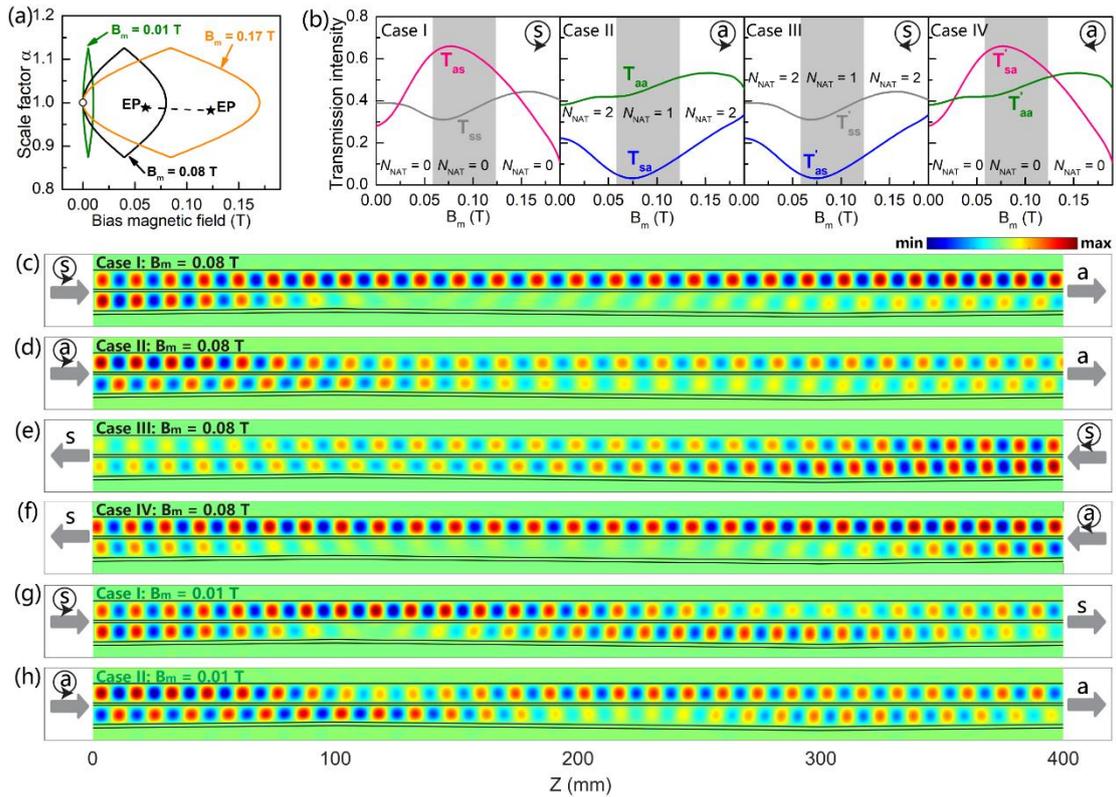

FIG. 3. (a) Three loops in parameter space, generated as examples with $B_m = 0.01$ T (green loop not enclosing any EP), 0.08 T (black loop enclosing one EP), and 0.17 T (yellow loop enclosing two EPs). The starting/end point lies at $B_0 = 0$ and $\alpha = 1$, corresponding to the symmetric phase. The black dashed line represents the broken phase line where the real parts of the eigenvalues coalesce. (b) Calculated transmission intensities for the four cases (see text for definition) as a function of $B_m$. The shaded region represents the field strengths where one EP is dynamically encircled, and a state flip occurs for cases I and IV only. Outside the shaded region, zero (left region) or two EPs (right region) are encircled. The number of non-adiabatic transitions (NATs) in the dynamical process, denoted by $N_{NAT}$, is given in different regions. (c)-(h) Numerically simulated $H_y$ field distributions in the waveguide system with different input modes and injection directions. The results with $B_m = 0.08$ T for cases I-IV are shown in (c)-(f), respectively, corresponding to an encircling of one EP. Panels (g) and (h) show results for cases I and II, respectively, with $B_m = 0.01$ T, corresponding to an encircling of zero EP. In all of the simulations, the frequency is 9 GHz and the system parameters are the same as those given in Fig. 1.

[42]. Our system serves as a controllable platform to study the dynamical process of state evolution on complex energy surfaces in non-Hermitian systems.

## III. STARTING/END POINT IN THE SYMMETRIC PHASE: CHIRAL DYNAMICS

We performed numerical simulations to demonstrate the effects arising from the dynamical encircling of EPs. We first consider encircling loops (Fig. 3(a)) with starting/end point in the symmetric phase where one eigenmode is symmetric and the other one antisymmetric. As the encircling can proceed either in the clockwise or the counter-clockwise direction and we can choose to excite either the symmetric or the antisymmetric mode at the starting point, there are four possible cases. Cases I and II correspond to counter-clockwise loops and cases III and IV clockwise loops. The injection is a symmetric mode for cases I and III and an antisymmetric mode for cases II and IV. We calculated the modal transmission intensities $T_{nm}$ ($T'_{nm}$), which are defined as the transmission from mode $m$ to mode $n$ in a counter-clockwise (clockwise) loop, where the subscript $s$ denotes the symmetric mode and $a$ the antisymmetric mode. The modal transmission intensities can reveal the behavior of mode switching. Figure 3(b) plots the calculated transmission intensities of the proposed system at 9 GHz as a function of $B_m$ for the four cases. In each plot, the left, middle (shaded), and right regions correspond to the bias field strengths at which zero, one, and two EPs are encircled respectively. We note that the system is still reciprocal (i.e., $T_{nm} = T'_{mn}$) in the presence of the transverse bias field since the cross section of the coupled waveguides has a mirror symmetry with respect to the plane $y = 0$ (see [43]; also see Supplemental Material for detailed descriptions of the mirror symmetry).

We first study the dynamics of encircling one EP for counter-clockwise loops. Case I in Fig. 3(b) shows that $T_{as} > T_{ss}$ in the shaded region (corresponding to one EP being encircled), so that the antisymmetric mode dominates the output. This means that a symmetric mode at the starting point ends up being an antisymmetric mode once the system has traveled one counter-clockwise loop in the parameter space. This phenomenon is representative of state flipping due to the self-intersecting Riemann energy surface in non-Hermitian systems. We note that the output is also an antisymmetric mode in case II, indicating that there is no state flip in this case. We take the loop generated at $B_m = 0.08$ T that encircles one EP (see Fig. 3(a)) as an example to explain the dynamics. The simulated field distributions ($H_y$ component) in the waveguide system for cases I and II are shown in Figs. 3(c) and 3(d), respectively. In Fig. 3(c), we see the mode switching, i.e., a symmetric mode excited at the left becomes an antisymmetric mode at the exit on the right. But mode switching is not observed in Fig. 3(d). To better understand the dynamics, we expand the transverse field distributions $\vec{f}(z)$ into a linear combination of the eigenfields (i.e., right eigenvectors) $\vec{r}_G(z)$ and $\vec{r}_L(z)$ of the configuration at a particular value of $z$. That is, we write $\vec{f}(z) = c_G \vec{r}_G(z) + c_L \vec{r}_L(z)$, where $c_G$ and $c_L$ are amplitudes, and the subscripts $G$ and $L$ are associated with the eigenmode with a lower loss (a relative 'gain' mode) and the eigenmode with a higher loss (a relative loss mode), respectively. The right eigenvectors $\vec{r}_G(z)$ and $\vec{r}_L(z)$ are typically not orthogonal since the system is non-Hermitian. We construct their corresponding left eigenvectors via $\vec{l}_{G(L)} = \vec{r}_{G(L)} - \langle \vec{r}_{G(L)} | \vec{r}_{L(G)} \rangle \vec{r}_{L(G)}$, and then determine the amplitudes by projecting the transverse field distributions onto the left eigenvectors (see Appendix A for details). The calculated amplitudes for cases I and II with $B_m = 0.08$ T are plotted in Figs. 4(a) and 4(b), respectively. We find that in case I the encircling process is stable and adiabatic since the state evolution takes place on the lower-loss Riemann sheet (also see the black line in Figs. 1(d) and 1(e)) so that $c_G$ dominates in the whole process. For case II, however, the state first propagates on the higher-loss Riemann sheet on which the state is known to be unstable [29-38]. There is a delay time [32] after which a NAT occurs (also see the black line in Figs. 1(f) and 1(g)), corresponding to the breakdown of adiabaticity [30-32]. After the NAT, the state propagates on the lower-loss Riemann sheet and no further NATs occur. As a result, the state returns to itself at the end of the loop because of the one NAT. The output for counter-clockwise loops is therefore always an antisymmetric mode, independent of the symmetry of the input mode, when one EP is encircled. By the same argument, the output for clockwise loops (i.e., cases III and IV) is always a symmetric mode (see Figs. 3(e) and 3(f); also see Supplemental Material for trajectories on the Riemann surface). This is the so-called chiral behavior of the transmission when one EP is encircled [29,30,32-38], i.e., the output depends solely on the encircling direction regardless of the symmetry of injection.

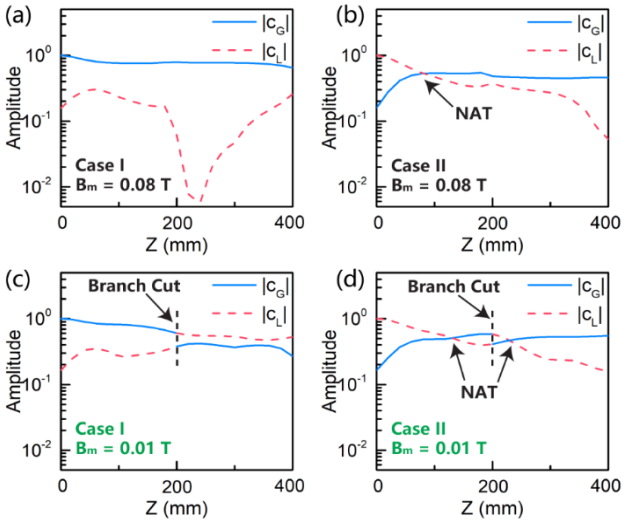

FIG. 4. Calculated amplitudes of the eigenstates along the waveguiding direction for (a) case I with $B_m = 0.08$ T, (b) case II with $B_m = 0.08$ T, (c) case I with $B_m = 0.01$ T, and (d) case II with $B_m = 0.01$ T, where $c_G$ and $c_L$ represent the coefficient of the eigenstate projected onto the lower-loss and higher-loss Riemann sheets, respectively. The black dashed lines in (c) and (d) show the existence of a branch cut via which the state can cross from one Riemann sheet to the other. The NAT is characterized by the crossing of two curves.

As we can vary the bias field strength to control the size of the loop and our system has two EPs, we can then study the dynamics when zero or two EPs are encircled. Figure 3(b) indicates that in the two non-shaded regions, the output mode is the same as the injection for all four cases as long as the loop is nowhere near the EP. The dynamics turns out to be rather complex. We take the loops generated at $B_m = 0.01$ T and $B_m = 0.17$ T as examples to investigate the dynamics. Figures 3(g) and 3(h) show respectively the $H_y$ field distributions in the waveguide system for cases I and II at $B_m = 0.01$ T. Although in both cases the state returns to itself after completing the loop, they exhibit different dynamics. To illustrate this point, we plot in Figs. 4(c) and 4(d) the corresponding amplitudes of the eigenmodes in the evolution process. The evolution process in case I is adiabatic so that the state returns to itself since the loop does not enclose any EP. In case II, however, the dynamics is highly non-adiabatic and two NATs occur throughout the process. As a result, the mode symmetry stays the same. The difference in the number of NATs can be understood intuitively by drawing the trajectories of the state evolution on the Riemann surface for cases I and II at $B_m = 0.17$ T, corresponding to an encircling of two EPs. Considering the topological structure of our system, encircling zero or two EPs should not make any difference to the behavior of mode switching because the chirality of one EP cancels the chirality of the other since they are derived from the same DP. The state acquires a geometric phase when two EPs are encircled, although this is unrelated to the symmetry of the output mode. We first consider case II (yellow lines in Figs. 1(f) and 1(g)). At the beginning, the state stays on the higher-loss Riemann sheet until the first NAT occurs, after which the state jumps to the lower-loss sheet on which it becomes stable. Later at $z = \sim 200$ mm, the state re-enters the higher-loss Riemann sheet via the branch cut (also see Fig. 4(d)) and becomes unstable again until the second NAT occurs. A total of two NATs occur in this highly non-adiabatic process. The evolution process in case I (yellow lines in Figs. 1(d) and 1(e)) is quite different since at first the state propagates on the lower-loss sheet. It is not until the state crosses over the branch cut (also see Fig. 4(c)) that it enters the higher-loss sheet. Interestingly, the expected NAT does not occur although in the rest of the process the state is not stable. This is because the delay time exceeds the time spent on the higher-loss sheet, indicating that the expected NAT may occur if we increase the length of the system (see Supplemental Material for a detailed discussion). The results of cases III and IV can be similarly understood (see Supplemental Material for trajectories on the Riemann surface). The number of NATs, denoted by $N_{NAT}$, is summarized in Fig. 3(b) for the four cases.

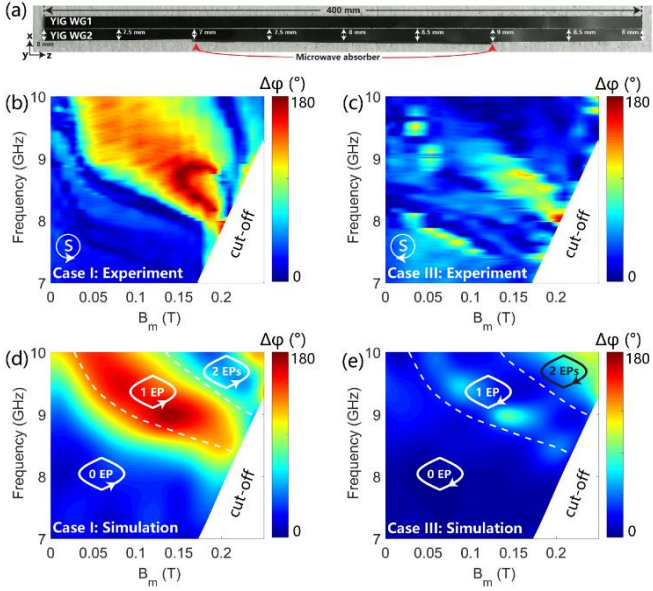
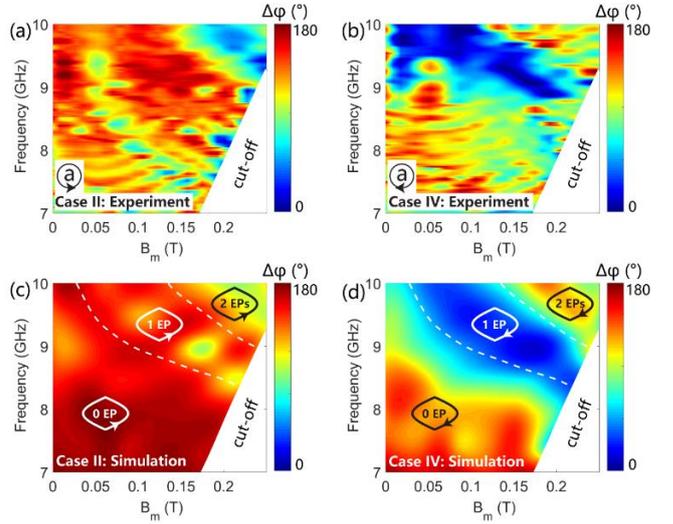

FIG. 5. (a) A photograph of the fabricated coupled YIG waveguides. Waveguide-1 measures $W \times H \times L = 8$ mm $\times$ 4 mm $\times$ 400 mm, while waveguide-2 measures $\alpha(z)W \times H \times L$ with the profile of $\alpha(z)$ shown in Fig. 2(d). The gap distance is $g = \sim 0.5$ mm. Microwave absorbers with the dimensions of $\sim 2$ mm $\times$ 1 mm $\times$ 200 mm are attached to the side of waveguide-2 to introduce loss. (b)-(c) Experimentally measured phase differences $\Delta\varphi$ at various bias fields $B_m$ and frequencies for case I (b) and case III (c). (d)-(e) Numerically simulated phase differences $\Delta\varphi$ as a function of the bias field $B_m$ and frequency for case I (d) and case III (e). The two dashed lines mark the calculated locations of EPs which partition the map into three regions depending on the number of EPs encircled. The phase difference was calculated based on the obtained transmission intensities (e.g., $\Delta\varphi = 2\arctan(T_{as}/T_{ss})$ for case I).

FIG. 6. (a)-(b) Experimentally measured phase differences $\Delta\varphi$ at various bias fields $B_m$ and frequencies for case II (a) and case IV (b). (c)-(d) Numerically simulated phase differences $\Delta\varphi$ as a function of the bias field $B_m$ and frequency for case II (c) and case IV (d).

We performed microwave experiments to demonstrate the above effects. A photograph of the fabricated samples is shown in Fig. 5(a) (see the figure caption for detailed parameters). The YIG waveguides were made from pure YIG with a saturation magnetization of $4\pi M_s \approx 1884$ G (produced by Nanjing Bi'ao Electronic Technology Co., Ltd.). The YIG waveguide-2 was created from a larger sample using a hand polishing machine and followed the shape designed in Fig. 2(d). The microwave absorber is attached to only half of YIG waveguide-2. This has been shown to be an effective way to minimize the dissipation of the system while keeping the topology of the system intact (see Ref. [35]; also see Supplemental Material for a discussion on the performance of such a system). We consider the phase difference $\Delta\varphi = |\varphi_1 - \varphi_2|$ as the criterion to determine the symmetry of the output mode, where $\varphi_1$ ($\varphi_2$) is the phase measured at the output side of waveguide-1 (waveguide-2). By definition, $\Delta\varphi = 0°$ corresponds to a symmetric mode whereas $\Delta\varphi = 180°$ an antisymmetric mode. In the experiment, the symmetric injection was excited using an $\sim 20$ mm long antenna, while the antisymmetric injection was excited using two $\sim 8$ mm long antennas which were connected to the source via a one-to-two power splitter and placed along opposite directions so that their currents were oscillating out of phase. An antenna $\sim 8$ mm in length was placed at the exit of waveguide-1 and waveguide-2 to detect their corresponding phases $\varphi_1$ and $\varphi_2$. All of the antennas were connected to an Agilent Technologies 8720ES Network Analyzer to record the transmission intensity and phase.

The measured phase differences as a function of the external field strength ($B_m$) and frequency are shown in Figs. 5(b) and 5(c), respectively, for cases I and III in which a symmetric mode is injected. We note in Fig. 5(b) that for each frequency above $\sim 8$ GHz there is a specific range of $B_m$ (in red) within which the system exhibits a state flip. This specific range shifts towards larger bias fields for lower frequencies. Figure 5(d) shows numerical simulation results for case I, which agrees well with the measurement. In the simulation, the relative permittivity of the absorber is set to $3+10i$ which can best match the experimental results. We also determine for each frequency the location of the EPs in the parameter space and mark them with the two white dashed lines in Fig. 5(d). The whole map is partitioned with these EP trajectories into three regions depending on the number of EPs encircled. The variation in the output mode symmetry with increasing bias field indeed reflects a change in the number of EPs encircled in the parameter space. In contrast, the output in case III is always a symmetric mode regardless of the number of EPs encircled (Figs. 5(c) and 5(e)). This thus demonstrates experimentally the breakdown of adiabaticity. The deviation between experimental and numerical results comes from the imperfectness of sample which is made by hand polishing. In addition, the input mode is excited using antennas placed outside the waveguide, and as such, its symmetry can only be approximately correct. However, even by just comparing the experimental results themselves (Figs. 5(b) and 5(c)), there is obviously a marked difference for the case of encircling one EP. The phase differences for cases II and IV injected with an antisymmetric mode are shown in Fig. 6. All these results are consistent with the analysis in Fig. 3, convincingly demonstrating the behavior of mode switching when different numbers of EPs are encircled, i.e., a chiral behavior is found when one EP is encircled and no state flip occurs when zero or two EPs are encircled. Results of a control experiment are given in Supplemental Material. The chiral nature of the dynamics of encircling one EP has been exploited for asymmetric mode switching [35,37]. Since the external field in this work can be tuned continuously, our system can be applied to the switching of modes controlled with external fields, i.e., manipulating the symmetry of the output state by dynamically encircling different numbers of EPs. Note that the microwave absorber in our design is attached on the YIG waveguide-2 with a varying width. We can also attach the absorber on the straight YIG waveguide-1 and the physics is the same.

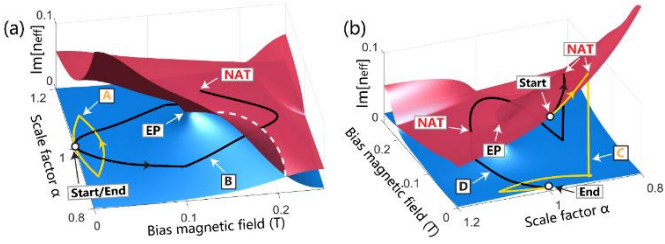

FIG. 7. (a) Calculated imaginary part of the effective mode index as a function of the bias field and scale factor of the system at 11.5 GHz with structure parameters: $W = 8$ mm, $H = 4$ mm, $g = 0.5$ mm, $w = 2$ mm, and $h = 3$ mm. The relative permittivity of the absorber is set to $4+15i$. The yellow and black lines mark the state evolution trajectory for configurations A and B (see text for definition), respectively, and the white dashed line marks the branch cut. (b) Same as those in (a) except that the trajectories are for configurations C and D.

## IV. STARTING/END POINT IN THE BROKEN PHASE: NON-CHIRAL DYNAMICS

In the previous section, we have explored the dynamical behavior when zero, one, or two EPs are dynamically encircled. The starting/end point lay in the symmetric phase, which is also the configuration explored in all previous works [34-38]. In this section, we show that when the starting/end point moves to the broken phase, the dynamical encircling would result in a non-chiral transmission behavior, in stark contrast to the chiral behavior when the system starts from a point in the symmetric phase.

We first describe the principle behind the system design. The starting/end point is still fixed at $B_0 = 0$ and $\alpha = 1$ for ease of experimental realization. To fulfil this requirement, the DP in the lossless system should be located close enough to the zero-bias field. We find in Fig. 1(b) that higher frequencies meet this requirement so we set the frequency to 11.5 GHz and choose the following system parameters: $W = 8$ mm, $H = 4$ mm, and $g = 0.5$ mm. The DP is then located at $B_0 = 0.047$ T (black circle in Fig. 1(b)), which is also approximately the center of the broken phase when microwave absorbers are attached [42]. We should choose a stronger absorber to ensure that the lossy system stays in the broken phase region at $B_0 = 0$ and $\alpha = 1$. To verify the design concept, we calculated the effective mode index of the system with a stronger absorber ($w = 2$ mm, $h = 3$ mm, $\varepsilon = 4+15i$) and show the Riemann surface in Fig. 7(a) (imaginary part). There is a large gap between the two Riemann sheets at $B_0 = 0$, confirming that the starting/end point indeed lies in the broken phase, where one eigenmode is nearly lossless (see the blue sheet) and the other one more lossy (see the red sheet). This is a result of symmetry breaking, i.e., the power flow of the lossless/lossy mode mainly propagates in the lossless/lossy YIG waveguide. As expected, there is only one EP which lies at $B_0 = 0.121$ T and $\alpha = 1.02$ (also see the parameter space in Fig. 8(a)).

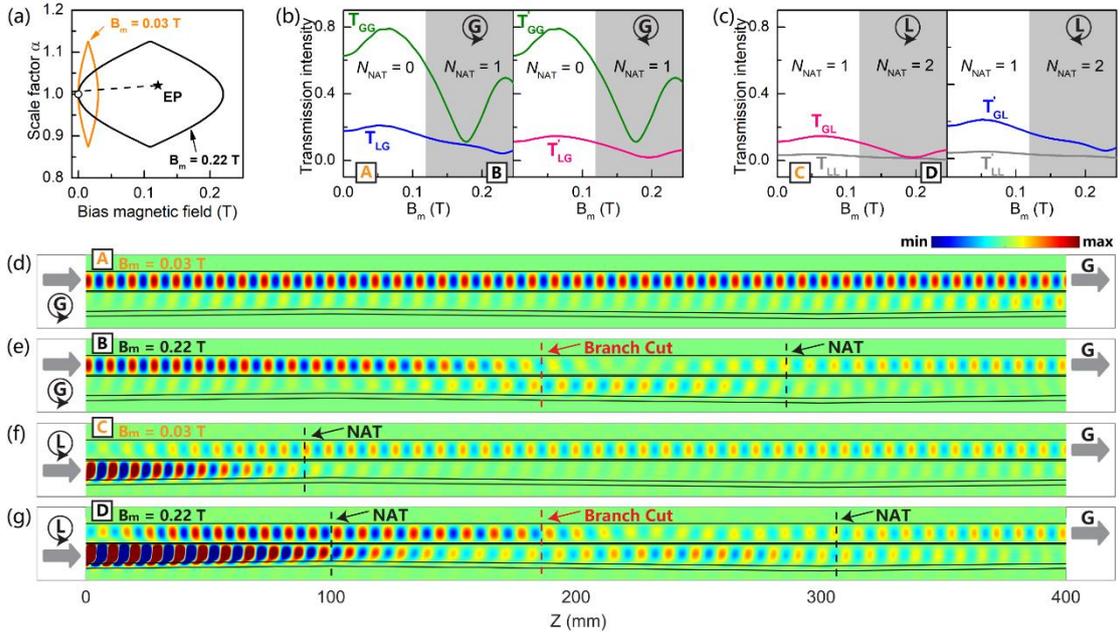

FIG. 8. (a) Loops in the parameter space generated with $B_m = 0.03$ T (yellow loop not enclosing any EP) and 0.22 T (black loop enclosing one EP). The starting/end point lies at $B_0 = 0$ and $\alpha = 1$, corresponding to the broken phase. The black dashed line represents the broken phase line where the real parts of the eigenvalues coalesce. (b) Calculated transmission intensities as a function of $B_m$ for counter-clockwise loops and clockwise loops with a 'gain' mode as the injection. The shaded region represents the area where one EP is dynamically encircled. The number of NATs in the dynamical process is given in different regions. (c) Same as those in (b) except that the injection is a loss mode. (d)-(g) Numerically simulated $H_y$ field distributions in the waveguide system for configurations A-D (see text for definition). The black dashed lines and red dashed lines mark the NAT and branch cut, respectively. System parameters are the same as those given in Fig. 7.

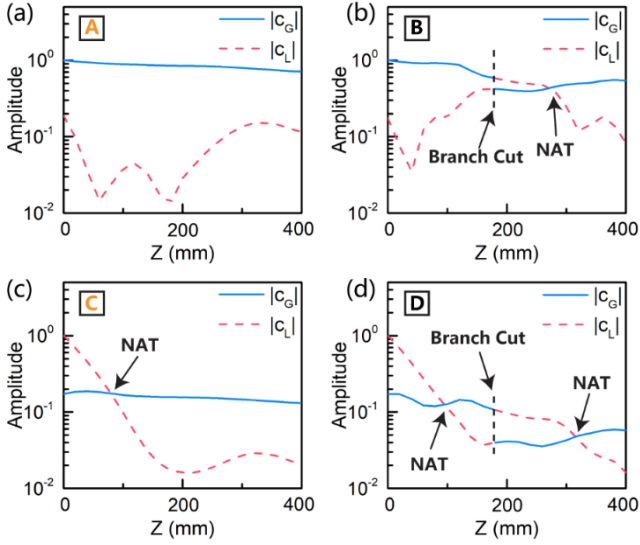

FIG. 9. (a)-(d) Calculated amplitudes of the eigenstates along the waveguiding direction for configurations A-D.

The transmission intensities of the proposed system (see Figs. 2(a) and 2(b) for the schematic diagram) with the parameters mentioned above are calculated as a function of $B_m$ to investigate the behavior when the EP is encircled with the starting/end point in the broken phase. The transmission intensity $T_{nm}$ ($T'_{nm}$) is defined in the same way as that in Fig. 3(b), except that here we use subscripts $G$ and $L$ to denote the nearly lossless (i.e., a relative 'gain') mode and the lossy mode, respectively. The results with a 'gain' mode injection and a loss mode injection are plotted in Figs. 8(b) and 8(c), respectively, in which the region where one EP is encircled is shaded. Under each injection, the results of counter-clockwise loops and clockwise loops look almost the same, indicating a non-chiral transmission behavior which is distinct from the chiral behavior found when the starting/end point is in the symmetric phase (see Fig. 3(b)). More interestingly, we find that the output is always a 'gain' mode, regardless of the details such as the input, encircling direction, or even the number of EPs encircled. To investigate the underlying physics, we study four configurations in this section. Configurations A and B are counter-clockwise loops generated at $B_m = 0.03$ T and 0.22 T, corresponding to an encircling of zero and one EP, respectively, with a 'gain' mode as the injection (see Fig. 8(b)). It is the same for configurations C and D but with a loss mode as the input (see Fig. 8(c)).

Figures 8(d)-8(g) show the $H_y$ field profiles in the waveguide system for configurations A-D, and the amplitudes of their eigenmodes extracted from the field profiles are plotted in Figs. 9(a)-9(d), respectively. We first analyze the small encircling loop that excludes the EP. Configuration A is the simplest case in the sense that the state evolution stays all the time on the lower-loss Riemann sheet (see the yellow line in Fig. 7(a)). As a result, the dynamical process is stable and adiabatic (Fig. 9(a)), as verified by the calculated results showing a concentration of power flow in YIG waveguide-1 in the whole process (Fig. 8(d)). Configuration C is different in that a loss mode is injected. The process is unstable at first until a NAT to the lower-loss Riemann sheet occurs, and the state becomes stable for the rest of the process (see the yellow line in Fig. 7(b)). This NAT can be seen from the field profiles in Fig. 8(f). It is characterized by a power transfer from waveguide-2 to waveguide-1 (see the black dashed line; also refer to Fig. 9(c)).

Configurations B and D in which one EP is encircled exhibit rather complex dynamics. Configuration B starts with a stable evolution process. As the state encircles the EP, it enters the higher-loss Riemann sheet via the branch cut. A NAT then occurs after some delay time, causing the state to jump to the lower-loss sheet, after which the stable state arrives at the end point as a 'gain' mode. The trajectory of this process is plotted with a black line in Fig. 7(a), according to the simulated field profiles in Fig. 8(e) and amplitudes of the eigenmodes in Fig. 9(b). The branch cut is characterized by a power transfer from waveguide-1 to waveguide-2 (see the red dashed line in Fig. 8(e)). Configuration D has the most complex dynamics. The state is unstable at first so that it jumps to the lower-loss sheet via a NAT. The following process is the same as that of configuration B, i.e., the state re-enters the higher-loss sheet via the branch cut, experiences a second NAT and reaches the end point as a 'gain' mode (see the black line in Fig. 7(b); also see Figs. 8(g) and 9(d)).

The number of NATs obtained from the above analysis is summarized in Figs. 8(b) and 8(c), which shed light on the complex transmission behavior. When a 'gain' mode is injected, configuration A exhibits the highest transmission intensity since the state evolution is always on the lower-loss sheet. As the bias field is increased to enlarge the encircling loop, the EP can be encircled. The state is then able to climb up to the higher-loss sheet so that the transmission drops considerably. The delay time of the unstable state on the higher-loss sheet is determined by the system parameters, especially the absorber properties. The transmission dip in Fig. 8(b) ($\sim B_m = 0.17$ T) can thus be interpreted as a process featuring the largest energy attenuation considering both the encircling loop and the delay time. It is also evident that the transmission intensity should be much lower when a loss mode is injected, e.g., configurations C and D.

We performed experiments to verify the above analysis. In the experiments, the 'gain' mode and loss mode were excited by putting an $\sim 8$ mm antenna near the entrance of waveguide-1 and waveguide-2, respectively. The measured transmission spectra at 11.5 GHz are shown in Figs. 10(a) and 10(b), which agree well with the numerical results in Figs. 8(b) and 8(c), confirming the non-chiral transmission behavior. We also measured the electric field intensity to elucidate the NATs in the dynamical process. In the experimental measurement, we put an ~8 mm long antenna on top of each YIG waveguide to measure their corresponding electric field intensity as a function of $z$. The measured results of counter-clockwise loops with a 'gain' injection at different $B_m$ values are shown in Figs. 10(c) and 10(d), respectively, for waveguide-1 and waveguide-2. We find in Fig. 10(d) that the field intensity in waveguide-2 is very weak at $z = 0$. In the range $B_m > \sim 0.125$ T, there is a considerable increase in the field intensity at the center of the system ($z = \sim 20$ mm). This is a typical feature of the branch cut (see the dashed ellipse) and confirms the dynamical encircling of one EP in experiment. The state then climbs up to the higher-loss Riemann sheet via the branch cut so that it becomes unstable allowing a NAT to occur, as shown by the drastic decrease in the field intensity at $z = \sim 30$ mm (see the dashed ellipse). The number of NATs is therefore a good indicator of the number of EPs encircled. The same measurements but with a loss injection are shown in Figs. 10(e) and 10(f). The first NAT appears at $z = \sim 5$ mm for all values of $B_m$ (see the white dashed ellipse in Fig. 10(f)), after which the state jumps to the lower-loss Riemann sheet associated with a sudden increase in the field intensity in waveguide-1 (see Fig. 10(e)). The following dynamics is the same as that with a 'gain' injection, i.e., the state re-enters the higher-loss sheet via the branch cut and experiences a second NAT (see the two dashed ellipses in Fig. 10(f)), for the loops enclosing one EP only. The experimentally measured transmission spectra and number of NATs extracted from the field profiles strongly support the numerical simulations and demonstrate the non-chiral behavior when the starting/end point lies in the broken phase.

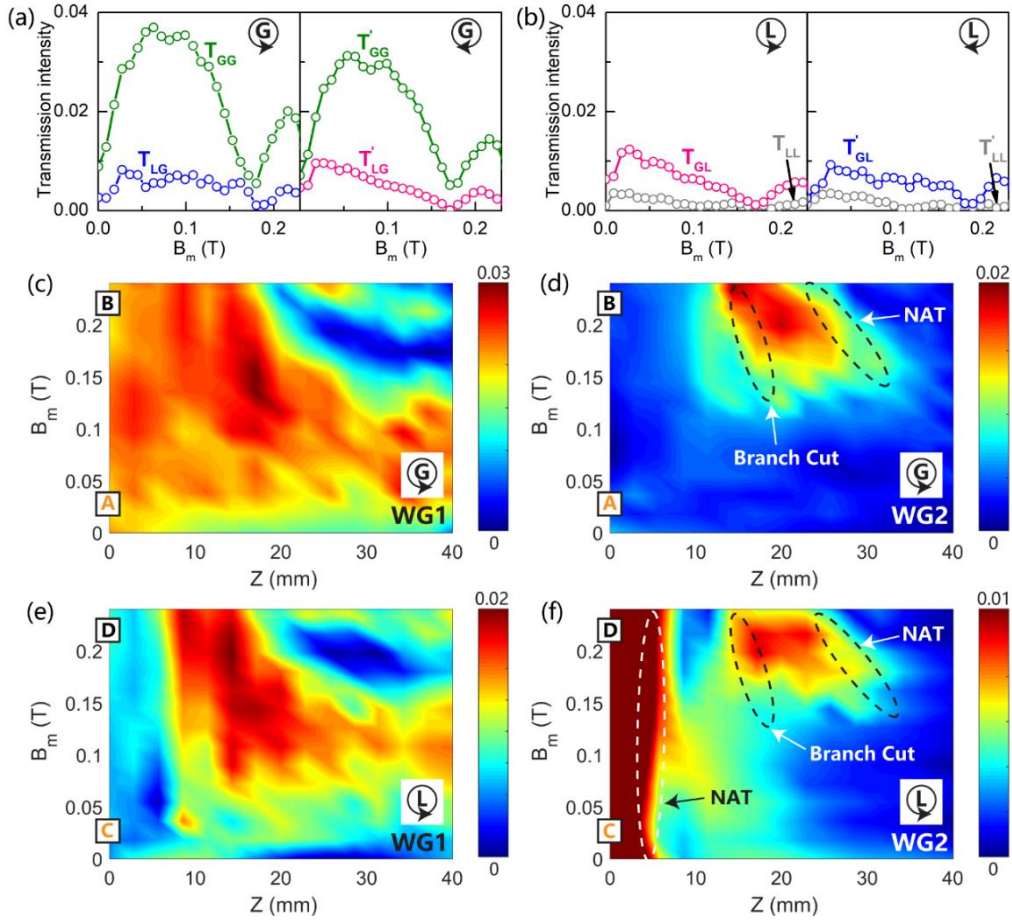

FIG. 10. (a)-(b) Experimentally measured transmission intensities at 11.5 GHz as a function of $B_m$ with a 'gain' mode (a) or a loss mode (b) as the injection. The system parameters are $W = 8$ mm, $H = 4$ mm, and $g = 0.5$ mm. A microwave absorber stronger than the one used in Fig. 5(a) with the dimensions of ~3 mm × 2 mm × 400 mm is attached to waveguide-2. (c)-(f) Experimentally measured surface electric field intensities along the waveguiding direction at 11.5 GHz for different values of $B_m$. Results for counter-clockwise loops with a 'gain' mode as the injection are shown in (c) and (d) for waveguide-1 (WG1) and waveguide-2 (WG2), respectively, while results for counter-clockwise loops with a loss injection are shown in (e) and (f).

## V. THEORETICAL DEMONSTRATION OF NON-CHIRAL DYNAMICS

In this section, we consider the time evolution of a simple non-Hermitian Hamiltonian to show that the dynamics is non-chiral when the starting point lies in the broken phase. We consider a two-state system governed by $i\partial_t |\psi(t)\rangle = H(t)|\psi(t)\rangle$, where the generic time-dependent Hamiltonian has the form

$$H(t) = \begin{pmatrix} ig(t) + \delta(t) & \kappa \\ \kappa & -ig(t) - \delta(t) \end{pmatrix}, \quad (1)$$

and $|\psi(t)\rangle = [a(t), b(t)]^T$ is the state vector at time $t$. It is easy to see that $g(t)$ and $\delta(t)$ represent respectively the amount of gain/loss and detuning, and the coupling strength is denoted by $\kappa$ which for simplicity is set to be -1. We use this simple Hamiltonian to highlight the fact that the phenomenon we have observed is rather generic, and not just specific to our particular experimental configuration. A two-parameter space with $g$ and $\delta$ is shown in Fig. 11(a), where we have a pair of EPs at $g = \pm 1$ and $\delta = 0$. The red line and green line correspond to the broken phase and symmetric phase, respectively. We consider an encircling loop parameterized by

$$g(t) = 1 - \rho\cos(\gamma t), \quad \delta(t) = \rho\sin(\gamma t), \quad (2)$$

where $\rho$ denotes the radius of the loop (see Fig. 11(a)), and $\gamma$ is a measure of adiabaticity. A positive $\gamma$ leads to a counter-clockwise loop whereas a negative $\gamma$ a clockwise loop. The starting point and end point are chosen at $t_0 = -\pi/|\gamma|$ and $t_{end} = \pi/|\gamma|$, respectively, so that they both lie in the broken phase. There are two eigenmodes, i.e., a gain mode and a loss mode, at the starting/end point. The corresponding eigenvalues are $\lambda_G = -i\sqrt{\rho^2 + 2\rho}$ and $\lambda_L = i\sqrt{\rho^2 + 2\rho}$, while the right eigenvectors are $|\psi_G\rangle = [1, i(\rho + 1 - \sqrt{\rho^2 + 2\rho})]^T$ and $|\psi_L\rangle = [1, i(\rho + 1 + \sqrt{\rho^2 + 2\rho})]^T$.

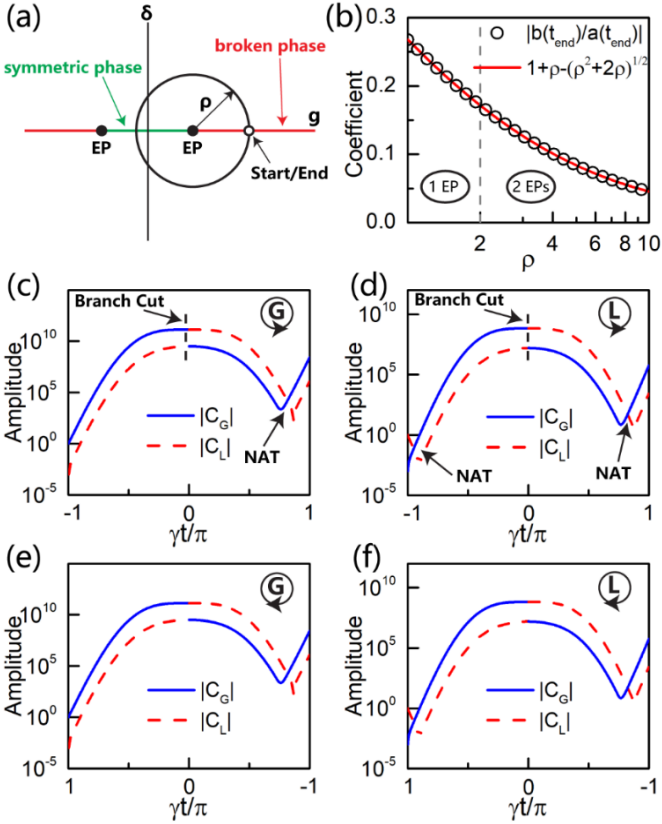

FIG. 11. (a) The $g$-$\delta$ parameter space where a pair of EPs locates at $g = \pm 1$ and $\delta = 0$. The circle with a radius $\rho$ depicts a trajectory to encircle the EP with the starting point in the broken phase. The red line and green line denote the broken phase and symmetric phase, respectively. (b) Calculated $|b(t_{end})/a(t_{end})|$ as a function of $\rho$ with $\gamma = 0.1$ (counter-clockwise loops) and $-0.1$ (clockwise loops). The region with $\rho < 2$ corresponds to the dynamical encircling of one EP whereas that with $\rho > 2$ corresponds to the encircling of two EPs. We performed four calculations (i.e., counter-clockwise/clockwise loop with a gain/loss input) and the results are all the same as shown by the black circles. The red line shows the value of $\rho + 1 - \sqrt{\rho^2 + 2\rho}$ as a function of $\rho$, which matches well with the black circles, indicating that the final state is always a gain state. (c)-(f) Calculated amplitudes of the eigenvectors for (c) counter-clockwise loop with a gain input, (d) counter-clockwise loop with a loss input, (e) clockwise loop with a gain input, and (f) clockwise loop with a loss input. In the calculations, we choose $\rho = 1$ and $\gamma = \pm 0.1$, corresponding to the dynamical encircling of an EP.

We first calculated the evolution of the state vector $|\psi(t)\rangle$ by numerically solving the time-dependent equation. The state vector at each time step can be decomposed as a sum of the instantaneous right eigenvectors, i.e., $|\psi(t)\rangle = C_G|\psi_G(t)\rangle + C_L|\psi_L(t)\rangle$, where $|\psi_G(t)\rangle$ and $|\psi_L(t)\rangle$ are instantaneous right eigenvectors that can be solved from the instantaneous Hamiltonian, and their corresponding amplitudes $C_G$ and $C_L$ can be obtained by projecting the state vector onto the left eigenvectors. This process is exactly the same as that for calculating the amplitudes of the instantaneous eigenmodes in the coupled waveguide system (see Figs. 4 and 9). The amplitudes of the instantaneous eigenvectors with $\rho = 1$ and $\gamma = \pm 0.1$, corresponding to the dynamical encircling of an EP, are shown in Figs. 11(c)-11(f) for different input modes and encircling directions as indicated in the figures. The blue lines are associated with the gain eigenstate while the red dashed lines the loss eigenstate. We can infer from the results that the output is always dominated by the gain eigenstate, regardless of the input state and encircling direction. For any input state, the dynamics for counter-clockwise and clockwise loops are nearly the same. There is one NAT when a gain state is injected while two NATs with a loss state injection. The results of this simple model well reproduce the features of the coupled waveguide system (see Figs. 9(b) and 9(d)). A more rigorous way to identify the output state is to calculate $|b(t_{end})/a(t_{end})|$ as a function of $\rho$. We find no matter which state is injected and which direction the encircling takes, the results are the same as shown by the black circles in Fig. 11(b) where we fix $\gamma = \pm 0.1$. We know the gain state has the eigenvector $|\psi_G\rangle = [1, i(\rho + 1 - \sqrt{\rho^2 + 2\rho})]^T$ so that the corresponding ratio $|b/a| = \rho + 1 - \sqrt{\rho^2 + 2\rho}$. This expression is plotted as a function of $\rho$ in Fig. 11(b) by the red line which coincides with $|b(t_{end})/a(t_{end})|$, indicating that the final state is always a gain state when the starting point lies in the broken phase, no matter whether one or two EPs are encircled (i.e., $\rho < 2$ for encircling one EP and $\rho > 2$ for two EPs).

In fact, this preferred final state and the corresponding non-chiral dynamics can be proved mathematically by deriving an analytical form of $|b(t_{end})/a(t_{end})|$. The above model Hamiltonian and trajectory in the parameter space have been analyzed recently [34], where the authors studied the dynamics with the starting point in the symmetric phase and derived a closed-form expression of the state evolution. We adopt the same method to study our case. The key to the derivation is to first recast Eq. (1) into a second order differential equation for $a(t)$, e.g., $d^2a(t)/dt^2 - [\rho^2 e^{2i\gamma t} - \rho(2 + i\gamma)e^{i\gamma t}]a(t) = 0$, which can further be reduced to a degenerate hypergeometric differential equation. We first consider $\gamma > 0$ and the solution can be written as a sum of confluent hypergeometric functions of the first kind $F$ and second kind $U$. By applying initial conditions, the state vector can be expressed in the form of a transfer matrix

$$[a(t), b(t)]^T = \sigma(t) M_1(t) M_2 M_3 [a(t_0), b(t_0)]^T, (3)$$

where $\sigma(t) = i\Gamma(i/\gamma) e^{i(\rho/\gamma)(e^{i\gamma t}-1)}$ with $\Gamma$ being the gamma function and the matrices are

$$M_1(t) = \begin{bmatrix} F^{(0)} & U^{(0)} \\ iF^{(0)} + 2\rho e^{i\gamma t} F^{(1)}/\gamma & iU^{(0)} - 2\rho e^{i\gamma t} U^{(1)}/\gamma \end{bmatrix}, (4a)$$

$$M_2 = \begin{bmatrix} \rho U^{(0)}_{t=-\pi/\gamma}/\gamma + 2i\rho U^{(1)}_{t=-\pi/\gamma}/\gamma^2 & -U^{(0)}_{t=-\pi/\gamma} \\ -\rho F^{(0)}_{t=-\pi/\gamma}/\gamma + 2i\rho F^{(1)}_{t=-\pi/\gamma}/\gamma^2 & F^{(0)}_{t=-\pi/\gamma} \end{bmatrix}, (4b)$$

$$M_3 = \begin{bmatrix} 1 & 0 \\ (1+\rho)/\gamma & i/\gamma \end{bmatrix}, (4c)$$

where $F^{(n)}$ and $U^{(n)}$ represent confluent hypergeometric functions [44] $F(n+i/\gamma, n+1, -2i\rho e^{i\gamma t}/\gamma)$ and $U(n+i/\gamma, n+1, -2i\rho e^{i\gamma t}/\gamma)$ respectively. The mathematical techniques used to solve the differential equation can be found in Ref. [34]. Our formulas are slightly different from those in Ref. [34] (see Eqs. (6a)-(6c) there) since here we have the initial condition $t_0 = -\pi/\gamma$ (i.e., starting point in the broken phase) whereas the starting point in Ref. [34] lies in the symmetric phase with $t_0 = 0$.

We now take a closer look at Eqs. (4a)-(4c). We focus on the final time step $t_{end} = \pi/\gamma$ and we introduce a matrix $M = M_1(t_{end}) M_2 M_3$ with matrix elements (see Appendix B for details)

$$m_{11} = -\frac{2\pi i}{\gamma \Gamma(i/\gamma)} F^{(0)}_{t=\pi/\gamma} F^{(0)}_{t=\pi/\gamma} + \frac{4\pi\rho}{\gamma^2 \Gamma(i/\gamma)} F^{(0)}_{t=\pi/\gamma} F^{(1)}_{t=\pi/\gamma}, (5a)$$
$$+ \frac{2i\rho}{\gamma^2} F^{(0)}_{t=\pi/\gamma} U^{(1)}_{t=\pi/\gamma} + \frac{2i\rho}{\gamma^2} F^{(1)}_{t=\pi/\gamma} U^{(0)}_{t=\pi/\gamma}$$

$$m_{12} = \frac{2\pi}{\gamma \Gamma(i/\gamma)} F^{(0)}_{t=\pi/\gamma} F^{(0)}_{t=\pi/\gamma}, (5b)$$

$$m_{21} = \frac{2\pi}{\gamma \Gamma(i/\gamma)} F^{(0)}_{t=\pi/\gamma} F^{(0)}_{t=\pi/\gamma} + \frac{8\pi\rho i}{\gamma^2 \Gamma(i/\gamma)} F^{(0)}_{t=\pi/\gamma} F^{(1)}_{t=\pi/\gamma}, (5c)$$
$$- \frac{8\pi\rho^2}{\gamma^3 \Gamma(i/\gamma)} F^{(1)}_{t=\pi/\gamma} F^{(1)}_{t=\pi/\gamma}$$

$$m_{22} = \frac{2\pi i}{\gamma \Gamma(i/\gamma)} F^{(0)}_{t=\pi/\gamma} F^{(0)}_{t=\pi/\gamma} - \frac{4\pi\rho}{\gamma^2 \Gamma(i/\gamma)} F^{(0)}_{t=\pi/\gamma} F^{(1)}_{t=\pi/\gamma}. (5d)$$
$$+ \frac{2i\rho}{\gamma^2} F^{(0)}_{t=\pi/\gamma} U^{(1)}_{t=\pi/\gamma} + \frac{2i\rho}{\gamma^2} F^{(1)}_{t=\pi/\gamma} U^{(0)}_{t=\pi/\gamma}$$

It is difficult to further simplify the above formulas but it is instructive to consider some limiting cases. Here we choose a finite $\gamma$ and let $\rho \to \infty$, corresponding to the dynamical encircling of two EPs. A big enough $\rho$ can make the system parameters change slowly enough so that it will not introduce non-adiabaticity into the system, which means the non-adiabaticity (if any) only comes from the non-Hermiticity induced by the gain and loss. In the limit $\rho \to \infty$, we have $z \to \infty$ for $F(p_1, p_2, z)$ and $U(p_1, p_2, z)$ since $z = 2i\rho/\gamma$. Then we can use the asymptotic expansions of $F(p_1, p_2, z) \approx (-z)^{-p_1} \Gamma(p_2)/\Gamma(p_2-p_1)$ and $U(p_1, p_2, z) \approx z^{-p_1}$ in the limit $z \to \infty$ (see Eqs. (4.1.3) and (4.1.12) in Ref. [44]), which leads to

$$F^{(0)}_{t=\pi/\gamma} \approx i\Gamma(i/\gamma)(2i\rho/\gamma)^{-i/\gamma}/(2\pi), (6a)$$
$$F^{(1)}_{t=\pi/\gamma} \approx -\gamma\Gamma(i/\gamma)(2i\rho/\gamma)^{-i/\gamma}/(4\pi\rho), (6b)$$
$$U^{(0)}_{t=\pi/\gamma} \approx (2i\rho/\gamma)^{-i/\gamma}, (6c)$$
$$U^{(1)}_{t=\pi/\gamma} \approx -i\gamma(2i\rho/\gamma)^{-i/\gamma}/(2\rho). (6d)$$

Inserting these asymptotic forms into Eqs. (5a)-(5d) can help simplify the expressions of the matrix elements. We find $m_{11} = m_{21} = m_{22} = 0$ and only $m_{12} \neq 0$ (see Appendix C for details). The final state $|b(t_{end})/a(t_{end})|$ then takes the form

$$|b(t_{end})/a(t_{end})| = \left| \frac{m_{21} a(t_0) + m_{22} b(t_0)}{m_{11} a(t_0) + m_{12} b(t_0)} \right| = \left| \frac{0}{m_{12} b(t_0)} \right| = 0. (7)$$

This analytic result demonstrates that no matter what state is injected, the final state always has $|b(t_{end})/a(t_{end})| \to 0$ when $\rho \to \infty$. Meanwhile, the ratio of the eigenvector element $|b/a|$ for the gain state (i.e., $a = 1, b = i(\rho + 1 - \sqrt{\rho^2 + 2\rho})$) and loss state (i.e., $a = 1, b = i(\rho + 1 + \sqrt{\rho^2 + 2\rho})$) is, respectively, 0 and $\infty$ in the limit $\rho \to \infty$. We can therefore conclude that the final state is always a gain mode for the generic model described by Eq. (1). The case of $\gamma < 0$ can also be proved to have $|b(t_{end})/a(t_{end})| \to 0$ using a similar process. This demonstrates the non-chiral dynamics when the starting point lies in the broken phase.

## VI. DISCUSSION ON THE ROLE OF THE STARTING/END POINT

As we have demonstrated chiral and non-chiral dynamics in Secs. III and IV, we discuss the role of the starting/end point in this section. The key to understanding the dynamics in the encircling process is the NAT, which may occur if there is more than one eigenstate in the non-Hermitian system and the predominant eigenstate is not the one with the lowest loss. The state in the dynamical process is stable only if it is on the Riemann sheet with the lowest loss. Once the state climbs up to a higher-loss Riemann sheet via the branch cut (e.g., see configurations B and D in Figs. 7-10), it becomes unstable but a NAT does not occur immediately. There is a certain system parameter-dependent delay before a NAT occurs, and this delay time plays a key role in the dynamical process. We have demonstrated both numerically and experimentally that the delay time can always be accessed in the systems studied in this work when one EP is encircled (see the state trajectories in Figs. 1 and 7). This fact implies that when the state approaches the end point, it would be on the lower-loss Riemann sheet (i.e., the blue sheet in Figs. 1 and 7), and the details of the previous dynamical process such as the injected mode and the number of NATs would all be forgotten by the system. As a result, the final state is solely determined by the encircling direction. We note in Fig. 1(d) that in the symmetric phase line, the blue sheet is discontinuous so that when the starting/end point lies there, counter-clockwise loops result in an antisymmetric output whereas clockwise loops a symmetric output, corresponding to a chiral transmission behavior. When the starting/end point moves to the broken phase where the blue sheet is continuous (see Fig. 7(a)), counter-clockwise loops and clockwise loops give the same output, i.e., the 'gain' mode, showing a non-chiral transmission behavior.

The chiral and non-chiral dynamics can also be understood using the theoretical model proposed in Sec. V. It was shown in Ref. [34] that when the encircling direction is reversed, the final state can be obtained by simply employing a transformation to the state vector $[a_{end}, b_{end}]^T \rightarrow [a_{end}^*, -b_{end}^*]^T$. When the starting/end point lies in the symmetric phase (i.e., $t = 0$), the eigenvectors are $|\psi_1\rangle = [1, e^{i\theta}]^T$ and $|\psi_2\rangle = [1, -e^{-i\theta}]^T$, where $\theta = \arcsin(1-\rho)$. It is easy to find $|\psi_1\rangle \rightarrow |\psi_2\rangle$ and $|\psi_2\rangle \rightarrow |\psi_1\rangle$ by doing the above transformation. The dynamics is chiral, i.e., changing the encircling direction flips the final state. The situation is quite different if the starting point is in the broken phase where the eigenvectors are $|\psi_G\rangle = [1, i(\rho+1-\sqrt{\rho^2+2\rho})]^T$ and $|\psi_L\rangle = [1, i(\rho+1+\sqrt{\rho^2+2\rho})]^T$. Performing the above transformation leads to $|\psi_G\rangle \rightarrow |\psi_G\rangle$ and $|\psi_L\rangle \rightarrow |\psi_L\rangle$, indicating that reversing the encircling direction does not affect the final state, which is exactly the non-chiral dynamics found in this work. This mathematical interpretation shows that the chiral and non-chiral dynamics are related to the properties of the eigenvectors in the symmetric and broken phase.

The above analysis actually applies to loops that enclose any number of EPs, provided that the NAT occurs each time when the state is on the higher-loss sheet. In fact, we have observed the non-chiral dynamics when zero EP (see Figs. 8(b) and 8(c)) and two EPs (see Fig. 11(b) and Eq. (7)) are dynamically encircled with the starting point in the broken phase. For the case with the starting point in the symmetric phase, we note in Fig. S3(a) (see Supplemental Material) that when the waveguide is longer ($L = 1000$nm), the dynamics is always chiral, independent of whether zero, one or two EPs are encircled. However, the chiral dynamics is not observed in our experimental system ($L = 400$nm) when zero and two EPs are encircled (see Figs. 5 and 6) which is due to the fact that our system is not long enough for the required NAT to occur. We note that a very recent paper [45] (with starting point in the symmetric phase) also pointed out that the chiral dynamics can be observed when the loop does not encircle any EP in the limit of very slow cycles, which is consistent with our analysis.

A natural question to ask is what the final state would be if the starting/end point lies somewhere far away from both the symmetric and broken phases. Although the above analysis indicates that the output is likely to be the mode with a lower loss, this is still an open question since the delay time is not always accessible. A stability loss delay was introduced in Ref. [32] to study the dynamical encircling of EPs and analytical form of the delay time for simple examples was derived. However, determining the delay time in realistic non-Hermitian systems remains a very complicated issue that needs further investigation.

## VII. CONCLUSION

In summary, we have shown both numerically and experimentally that a pair of ferromagnetic waveguides applied with non-uniform bias magnetic fields serves as a good platform to study dynamical processes in non-Hermitian systems. Such a system has two EPs and hence energy surfaces with a more complex topology. The trajectory of the state in the parameter space can be controlled *in situ*, as demonstrated experimentally. Using the proposed system, we have demonstrated experimentally the chiral dynamics when one EP is encircled. We can also dynamically encircle more than one EP experimentally to reveal the topological structure of the system possessing multiple EPs. More importantly, we revealed that whether the so-called chiral behavior can be observed depends on the location of the starting/end point of the encircling loop. When the starting/end point moves to the broken phase, the system exhibits non-chiral dynamics. We have proposed a theoretical model to interpret the underlying physics. Our results clarify the role of the starting/end point in the dynamical process of encircling EPs. The proposed system can be applied to mode switching controlled with an external parameter without changing or moving the sample. The platform can also be used to study more complex dynamics in non-Hermitian systems such as the encircling of high-order EPs.

## ACKNOWLEDGEMENTS

We thank Prof. Z.Q. Zhang and Dr. R.Y. Zhang for their valuable comments and suggestions. This work was supported by the Hong Kong Research Grants Council through grant no. AoE/P-02/12. X.-L.Z. was also supported by the National Natural Science Foundation of China (grant no. 61605056) and the China Postdoctoral Science Foundation (grant no. 2016M591480). B.H. was also supported by the National Natural Science Foundation of China (grant no. 11474212) and the visiting scholarship program for young scientists in Collaborative Innovation Center of Suzhou Nano Science and Technology and the Priority Academic Program Development (PAPD) of Jiangsu Higher Education Institutions.

## APPENDIX A: CONSTRUCTING LEFT EIGENVECTORS

There are two eigenmodes in the waveguide system propagating along the positive $z$-axis. Their transverse electric and magnetic fields are denoted by $\mathbf{E}_{\nu t}^R$, $\mathbf{E}_{\mu t}^R$ and $\mathbf{H}_{\nu t}^R$, $\mathbf{H}_{\mu t}^R$, where $\nu \neq \mu$ and the superscript $R$ indicates that they are right eigenvectors. The inner product of the two right eigenvectors in the waveguide configuration is defined as an integration over the entire waveguide cross section $S$:

$$\xi = \frac{1}{4}\int_S \left[\mathbf{E}_{vt}^R(x,y) \times \mathbf{H}_{\mu t}^{R*}(x,y) + \mathbf{E}_{\mu t}^{R*}(x,y) \times \mathbf{H}_{vt}^R(x,y)\right] \cdot \mathbf{z} ds. \quad (A1)$$

We have $\xi \neq 0$ since the system is non-Hermitian. The corresponding left eigenvector can then be constructed via

$$\begin{aligned} \mathbf{E}_{vt}^L &= \left(\mathbf{E}_{vt}^R - \xi \mathbf{E}_{\mu t}^R\right)/\zeta \\ \mathbf{H}_{vt}^L &= \left(\mathbf{H}_{vt}^R - \xi \mathbf{H}_{\mu t}^R\right)/\zeta \end{aligned}, \quad (A2)$$

where we have defined

$$\zeta = \int_S \frac{1}{2}\mathrm{Re}\left[\mathbf{E}_{vt}^R(x,y) \times \mathbf{H}_{vt}^{R*}(x,y)\right] \cdot \mathbf{z} ds - |\xi|^2. \quad (A3)$$

It is easy to verify that

$$\begin{aligned} \frac{1}{4}\int_S \left[\mathbf{E}_{vt}^L(x,y) \times \mathbf{H}_{\mu t}^{R*}(x,y) + \mathbf{E}_{\mu t}^{R*}(x,y) \times \mathbf{H}_{vt}^L(x,y)\right] \cdot \mathbf{z} ds &= 0, \\ \frac{1}{4}\int_S \left[\mathbf{E}_{vt}^L(x,y) \times \mathbf{H}_{vt}^{R*}(x,y) + \mathbf{E}_{vt}^{R*}(x,y) \times \mathbf{H}_{vt}^L(x,y)\right] \cdot \mathbf{z} ds &= 1 \end{aligned} \quad (A4)$$

which satisfies the orthogonal relation between left eigenvectors and right eigenvectors. Consider the transverse field distributions as a linear combination of the eigenfields:

$$\begin{aligned} \mathbf{E}_t(x,y) &= c_v \mathbf{E}_{vt}^R(x,y) + c_\mu \mathbf{E}_{\mu t}^R(x,y) \\ \mathbf{H}_t(x,y) &= c_v \mathbf{H}_{vt}^R(x,y) + c_\mu \mathbf{H}_{\mu t}^R(x,y) \end{aligned}. \quad (A5)$$

The amplitude coefficients can then be solved by projecting the transverse field distribution onto the left eigenvectors:

$$c_v = \frac{1}{4}\int_S \left[\mathbf{E}_{vt}^{L*}(x,y) \times \mathbf{H}_t(x,y) + \mathbf{E}_t(x,y) \times \mathbf{H}_{vt}^{L*}(x,y)\right] \cdot \mathbf{z} ds. \quad (A6)$$

In the simulations, we first performed full wave calculations to obtain all the field components in the system such as those in Figs. 8(d)-8(g). Then we performed eigenmode analysis at each position $z$ to get the right eigenvectors of a uniform waveguide of the same cross section. After that we constructed the left eigenvectors using Eq. (A2). Finally, we projected the transverse field distributions at each position $z$ onto the corresponding left eigenvectors using Eq. (A6) and we got the amplitudes of the eigenmodes which were then shown in, for example, Figs. 9(a)-9(d) to help understand the number of nonadiabatic transitions occurred in the process.

## APPENDIX B: DERIVATION OF EQ. 5

Starting from Eqs. (4a)-(4c), the elements of the matrix $M$ at the final time step $t_{\mathrm{end}} = \pi/\gamma$ are

$$\begin{aligned} m_{11} &= \frac{1}{\gamma}\left(U_{t=\pi/\gamma}^{(0)} F_{t=-\pi/\gamma}^{(0)} - F_{t=\pi/\gamma}^{(0)} U_{t=-\pi/\gamma}^{(0)}\right) + \frac{2i\rho}{\gamma^2}\left(F_{t=\pi/\gamma}^{(0)} U_{t=-\pi/\gamma}^{(1)} + U_{t=\pi/\gamma}^{(0)} F_{t=-\pi/\gamma}^{(1)}\right) \\ m_{12} &= \frac{i}{\gamma}\left(U_{t=\pi/\gamma}^{(0)} F_{t=-\pi/\gamma}^{(0)} - F_{t=\pi/\gamma}^{(0)} U_{t=-\pi/\gamma}^{(0)}\right) \\ m_{21} &= -\frac{2\rho}{\gamma^2}\left(F_{t=\pi/\gamma}^{(0)} U_{t=-\pi/\gamma}^{(1)} + U_{t=\pi/\gamma}^{(0)} F_{t=-\pi/\gamma}^{(1)}\right) - \frac{4i\rho^2}{\gamma^3}\left(F_{t=\pi/\gamma}^{(1)} U_{t=-\pi/\gamma}^{(1)} - U_{t=\pi/\gamma}^{(1)} F_{t=-\pi/\gamma}^{(1)}\right) \\ &\quad -\frac{i}{\gamma}\left(F_{t=\pi/\gamma}^{(0)} U_{t=-\pi/\gamma}^{(0)} - U_{t=\pi/\gamma}^{(0)} F_{t=-\pi/\gamma}^{(0)}\right) + \frac{2\rho}{\gamma^2}\left(F_{t=\pi/\gamma}^{(1)} U_{t=-\pi/\gamma}^{(0)} + U_{t=\pi/\gamma}^{(1)} F_{t=-\pi/\gamma}^{(0)}\right) \\ m_{22} &= \frac{1}{\gamma}\left(F_{t=\pi/\gamma}^{(0)} U_{t=-\pi/\gamma}^{(0)} - U_{t=\pi/\gamma}^{(0)} F_{t=-\pi/\gamma}^{(0)}\right) + \frac{2i\rho}{\gamma^2}\left(F_{t=\pi/\gamma}^{(1)} U_{t=-\pi/\gamma}^{(0)} + U_{t=\pi/\gamma}^{(1)} F_{t=-\pi/\gamma}^{(0)}\right) \end{aligned}. \quad (B1)$$

We use the properties of confluent hypergeometric functions to simplify these formulas. It is easy to find $F_{t=-\pi/\gamma}^{(0)} = F_{t=\pi/\gamma}^{(0)}$ and $F_{t=-\pi/\gamma}^{(1)} = F_{t=\pi/\gamma}^{(1)}$. On the other hand, the principal value of $U(p_1, p_2, z)$ is in the interval $-\pi < \arg(z) < \pi$. Apparently, $U_{t=-\pi/\gamma}^{(0)}$ and $U_{t=-\pi/\gamma}^{(1)}$ are out of this range so that we have to use a connection formula (see Eq. (2.2.20) in Ref. [44])

$$\begin{aligned} U_{t=-\pi/\gamma}^{(0)} &= \frac{2\pi i}{\Gamma(i/\gamma)} F_{t=\pi/\gamma}^{(0)} + U_{t=\pi/\gamma}^{(0)} \\ U_{t=-\pi/\gamma}^{(1)} &= -\frac{2\pi i}{\Gamma(i/\gamma)} F_{t=\pi/\gamma}^{(1)} + U_{t=\pi/\gamma}^{(1)} \end{aligned}. \quad (B2)$$

Inserting Eq. (B2) into Eq. (B1), we obtain more simplified expressions

$$\begin{aligned} m_{11} &= -\frac{2\pi i}{\gamma\Gamma(i/\gamma)} F_{t=\pi/\gamma}^{(0)} F_{t=\pi/\gamma}^{(0)} + \frac{4\pi\rho}{\gamma^2\Gamma(i/\gamma)} F_{t=\pi/\gamma}^{(0)} F_{t=\pi/\gamma}^{(1)} \\ &\quad + \frac{2i\rho}{\gamma^2} F_{t=\pi/\gamma}^{(0)} U_{t=\pi/\gamma}^{(1)} + \frac{2i\rho}{\gamma^2} F_{t=\pi/\gamma}^{(1)} U_{t=\pi/\gamma}^{(0)} \\ m_{12} &= \frac{2\pi}{\gamma\Gamma(i/\gamma)} F_{t=\pi/\gamma}^{(0)} F_{t=\pi/\gamma}^{(0)} \\ m_{21} &= \frac{2\pi}{\gamma\Gamma(i/\gamma)} F_{t=\pi/\gamma}^{(0)} F_{t=\pi/\gamma}^{(0)} + \frac{8\pi\rho i}{\gamma^2\Gamma(i/\gamma)} F_{t=\pi/\gamma}^{(0)} F_{t=\pi/\gamma}^{(1)} \\ &\quad - \frac{8\pi\rho^2}{\gamma^3\Gamma(i/\gamma)} F_{t=\pi/\gamma}^{(1)} F_{t=\pi/\gamma}^{(1)} \\ m_{22} &= \frac{2\pi i}{\gamma\Gamma(i/\gamma)} F_{t=\pi/\gamma}^{(0)} F_{t=\pi/\gamma}^{(0)} - \frac{4\pi\rho}{\gamma^2\Gamma(i/\gamma)} F_{t=\pi/\gamma}^{(0)} F_{t=\pi/\gamma}^{(1)} \\ &\quad + \frac{2i\rho}{\gamma^2} F_{t=\pi/\gamma}^{(0)} U_{t=\pi/\gamma}^{(1)} + \frac{2i\rho}{\gamma^2} F_{t=\pi/\gamma}^{(1)} U_{t=\pi/\gamma}^{(0)} \end{aligned}, \quad (B3)$$

which are exactly Eqs. (5a)-(5d) of the main text.

## APPENDIX C: DETERMINATION OF $m_{11}$, $m_{12}$, $m_{21}$ and $m_{22}$

Inserting Eqs. (6a)-(6d) into Eqs. (5a)-(5d), we have

$$m_{11} = -\frac{2\pi i}{\gamma \Gamma(i/\gamma)} F^{(0)}_{t=\pi/\gamma} F^{(0)}_{t=\pi/\gamma} + \frac{4\pi\rho}{\gamma^2 \Gamma(i/\gamma)} F^{(0)}_{t=\pi/\gamma} F^{(1)}_{t=\pi/\gamma}$$
$$+ \frac{2i\rho}{\gamma^2} F^{(0)}_{t=\pi/\gamma} U^{(1)}_{t=\pi/\gamma} + \frac{2i\rho}{\gamma^2} F^{(1)}_{t=\pi/\gamma} U^{(0)}_{t=\pi/\gamma}$$
$$= \frac{i\Gamma(i/\gamma)}{2\pi\gamma}(2i\rho/\gamma)^{-2i/\gamma} - \frac{i\Gamma(i/\gamma)}{2\pi\gamma}(2i\rho/\gamma)^{-2i/\gamma}$$
$$+ \frac{i\Gamma(i/\gamma)}{2\pi\gamma}(2i\rho/\gamma)^{-2i/\gamma} - \frac{i\Gamma(i/\gamma)}{2\pi\gamma}(2i\rho/\gamma)^{-2i/\gamma}$$
$$= 0 \quad , \text{(C1)}$$

$$m_{12} = \frac{2\pi}{\gamma \Gamma(i/\gamma)} F^{(0)}_{t=\pi/\gamma} F^{(0)}_{t=\pi/\gamma} = -\frac{\Gamma(i/\gamma)}{2\pi\gamma}(2i\rho/\gamma)^{-2i/\gamma} \text{, (C2)}$$

$$m_{21} = \frac{2\pi}{\gamma \Gamma(i/\gamma)} F^{(0)}_{t=\pi/\gamma} F^{(0)}_{t=\pi/\gamma} + \frac{8\pi\rho i}{\gamma^2 \Gamma(i/\gamma)} F^{(0)}_{t=\pi/\gamma} F^{(1)}_{t=\pi/\gamma} - \frac{8\pi\rho^2}{\gamma^3 \Gamma(i/\gamma)} F^{(1)}_{t=\pi/\gamma} F^{(1)}_{t=\pi/\gamma}$$
$$= -\frac{\Gamma(i/\gamma)}{2\pi\gamma}(2i\rho/\gamma)^{-2i/\gamma} + \frac{\Gamma(i/\gamma)}{\pi\gamma}(2i\rho/\gamma)^{-2i/\gamma} - \frac{\Gamma(i/\gamma)}{2\pi\gamma}(2i\rho/\gamma)^{-2i/\gamma}$$
$$= \Gamma(i/\gamma)(2i\rho/\gamma)^{-2i/\gamma}\left(-\frac{1}{2\pi\gamma} + \frac{1}{\pi\gamma} - \frac{1}{2\pi\gamma}\right)$$
$$= 0 \quad \text{, (C3)}$$

$$m_{22} = \frac{2\pi i}{\gamma \Gamma(i/\gamma)} F^{(0)}_{t=\pi/\gamma} F^{(0)}_{t=\pi/\gamma} - \frac{4\pi\rho}{\gamma^2 \Gamma(i/\gamma)} F^{(0)}_{t=\pi/\gamma} F^{(1)}_{t=\pi/\gamma}$$
$$+ \frac{2i\rho}{\gamma^2} F^{(0)}_{t=\pi/\gamma} U^{(1)}_{t=\pi/\gamma} + \frac{2i\rho}{\gamma^2} F^{(1)}_{t=\pi/\gamma} U^{(0)}_{t=\pi/\gamma}$$
$$= -\frac{i\Gamma(i/\gamma)}{2\pi\gamma}(2i\rho/\gamma)^{-2i/\gamma} + \frac{i\Gamma(i/\gamma)}{2\pi\gamma}(2i\rho/\gamma)^{-2i/\gamma} \quad . \text{(C4)}$$
$$+ \frac{i\Gamma(i/\gamma)}{2\pi\gamma}(2i\rho/\gamma)^{-2i/\gamma} - \frac{i\Gamma(i/\gamma)}{2\pi\gamma}(2i\rho/\gamma)^{-2i/\gamma}$$
$$= 0$$

Therefore, we see that $m_{11} = m_{21} = m_{22} = 0$ and $m_{12} \neq 0$ in the limit $\rho \to \infty$.